\documentclass[aps, prb, twocolumn, superscriptaddress, floatfix, showpacs]{revtex4}
\usepackage[english]{babel}
\usepackage[dvips]{graphicx}
\usepackage{psfrag}
\usepackage{amssymb}
\usepackage{subfigure}
\usepackage{amsmath}
\usepackage{float}

\begin{document}
\title{Effects of surface forces and phonon dissipation in a three-terminal nanorelay}

\author{L. M. Jonsson}
\email{mjonsson@fy.chalmers.se}
\affiliation{Department of Applied Physics, Chalmers University of Technology and G{\"o}teborg University, SE - 412 96 G{\"o}teborg, Sweden}

\author{T. Nord}
\affiliation{Department of Applied Physics, Chalmers University of Technology and G{\"o}teborg University, SE - 412 96 G{\"o}teborg, Sweden}

\author{S. Viefers}
\affiliation{Department of Applied Physics, Chalmers University of Technology and G{\"o}teborg University, SE - 412 96 G{\"o}teborg, Sweden}

\affiliation{Department of Physics, University of Oslo, 
             Postboks 1048 - Blindern, N-0316 Oslo, Norway }
\author{J. M. Kinaret}
\affiliation{Department of Applied Physics, Chalmers University of Technology and G{\"o}teborg University, SE - 412 96 G{\"o}teborg, Sweden}

\begin{abstract}
We have performed a theoretical analysis of the operational characteristics of 
a carbon-nanotube-based three-terminal nanorelay. We show that short range 
 and van der Waals forces have a significant impact on 
the characteristics of the relay and introduce design 
constraints. We also investigate the effects of dissipation 
due to phonon excitation in the drain contact, which 
changes the switching time scales of the system, decreasing the longest 
time scale by two orders of magnitude. We show 
that the nanorelay can be used as a memory element and investigate the 
dynamics and properties of such a device. 
\end{abstract}

\maketitle

\section{Introduction}
Nanoelectromechanical systems (NEMS) is a rapidly expanding research field with high potential for applications. \cite{roukes} Typically, the length scale of these systems is in the nanometer range and there is a strong coupling between the mechanical and electrical degrees of freedom. Key properties of NEMS are high natural frequencies, operation at ultra-low power, and low dissipation. Carbon nanotubes (CNTs) have extraordinary properties that make them ideal building blocks of NEMS. They are light, have very high Young's moduli \cite{wong} and can be elastically deformed without breaking. \cite{yakobsen} In addition, CNTs have a wide range of electric properties: they can for instance be either semi-conducting or metallic depending on detailed structure. \cite{dresselhaus} 

In this article, we analyze a nanorelay system in which a conducting multi-wall CNT (MWNT) is placed on a terrace in a silicon dioxide substrate and connected to three electrodes. This system has previously been studied by Kinaret \emph{et. al}, \cite{nanorelay} who introduced a simple model and discussed the characteristics of the device. In the present work, we focus on the effects of van der Waals and short range forces in the system. We show that the magnitudes of these forces are comparable to those of the elastic and electrostatic forces and that they need to be included in a more realistic model of the system. \cite{exjobb} A brief account on this topic has been presented in Ref. \onlinecite{ecs}. The surface forces alter the system's behavior, and the characteristics prove to be useful for an application of the relay as a memory element. 

We also investigate the dynamic performance of the proposed memory element and estimate its write times. We show that the switching time depend sensitively on dissipative surface processes related to tube-surface interactions when the tube impinges on the drain contact, and analyze this in more detail using a simple model phonon emission. The dissipation associated with this interaction changes the switching time scale of the relay, and consequently, change the typical write time of the proposed memory element.   

\section{The Model}
\begin{figure*}[t]
\centering
\psfrag{R_T}{\small{$R_{{\rm T}}$}}
\psfrag{C_G}[]{\small{$C_{{\rm g}}$}}
\psfrag{C_D}[]{\small{$C_{{\rm d}}$}}
\psfrag{V_G}{\small{$V_{{\rm g}}$}}
\psfrag{V_D}{\small{$V_{{\rm d}}$}}
\psfrag{V_S}{\small{$V_{{\rm s}}$}}
\psfrag{Z}{\small{$Z$}}
\psfrag{x}{\small{$x$}}
\psfrag{z}{\small{$z$}}
\psfrag{L}{\small{$L$}}
\psfrag{G}{\small{g}}
\psfrag{D}{\small{d}}
\psfrag{S}{\small{s}}
\psfrag{h}{\small{$\tilde{h}$}}
\subfigure[~] 
{
\label{fig:schematic}
\includegraphics[width=2.5in]{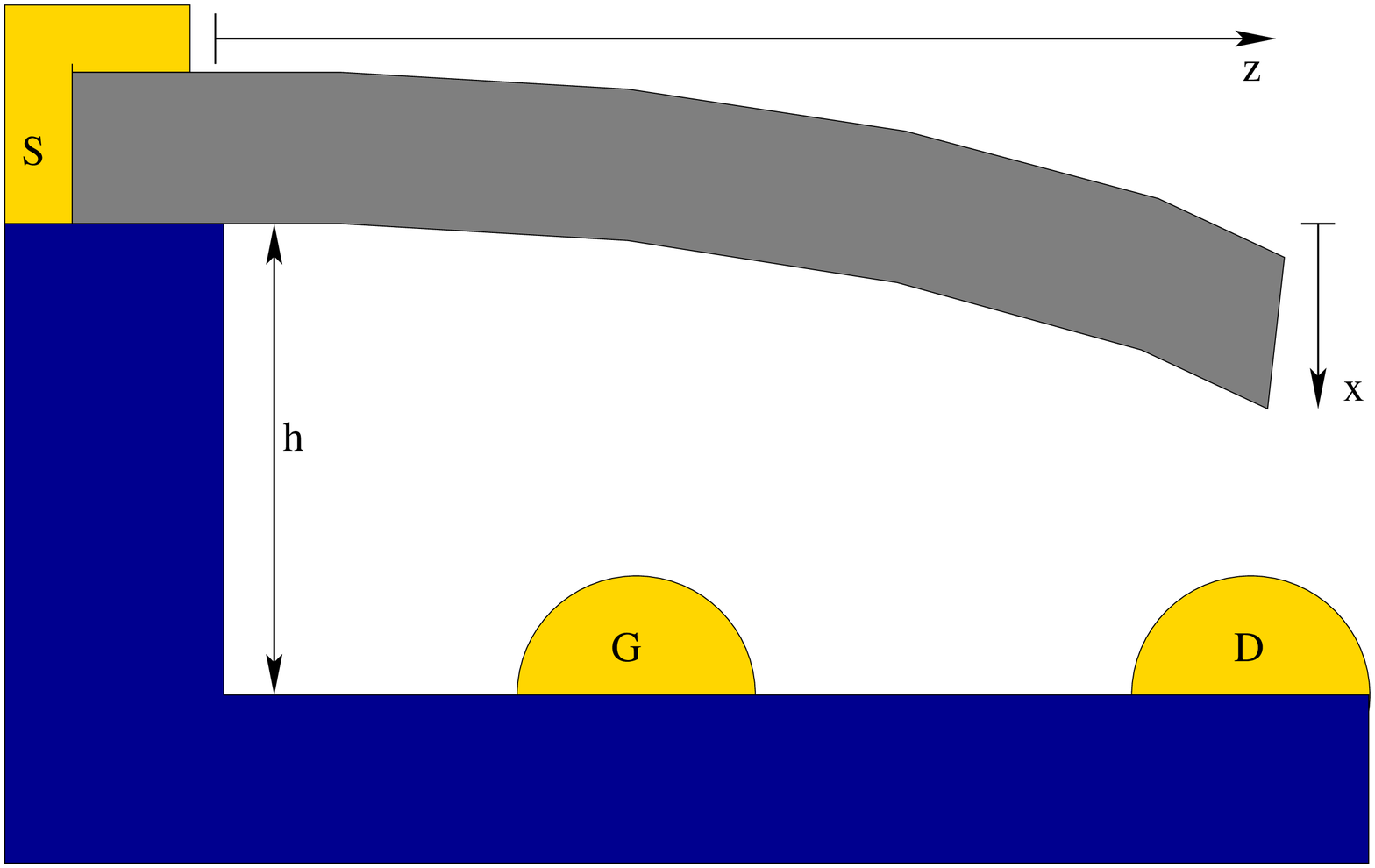}
}
\hspace{0cm}
\subfigure[~] 
{
\label{fig:eqvcircut}
\includegraphics[width=2.5in]{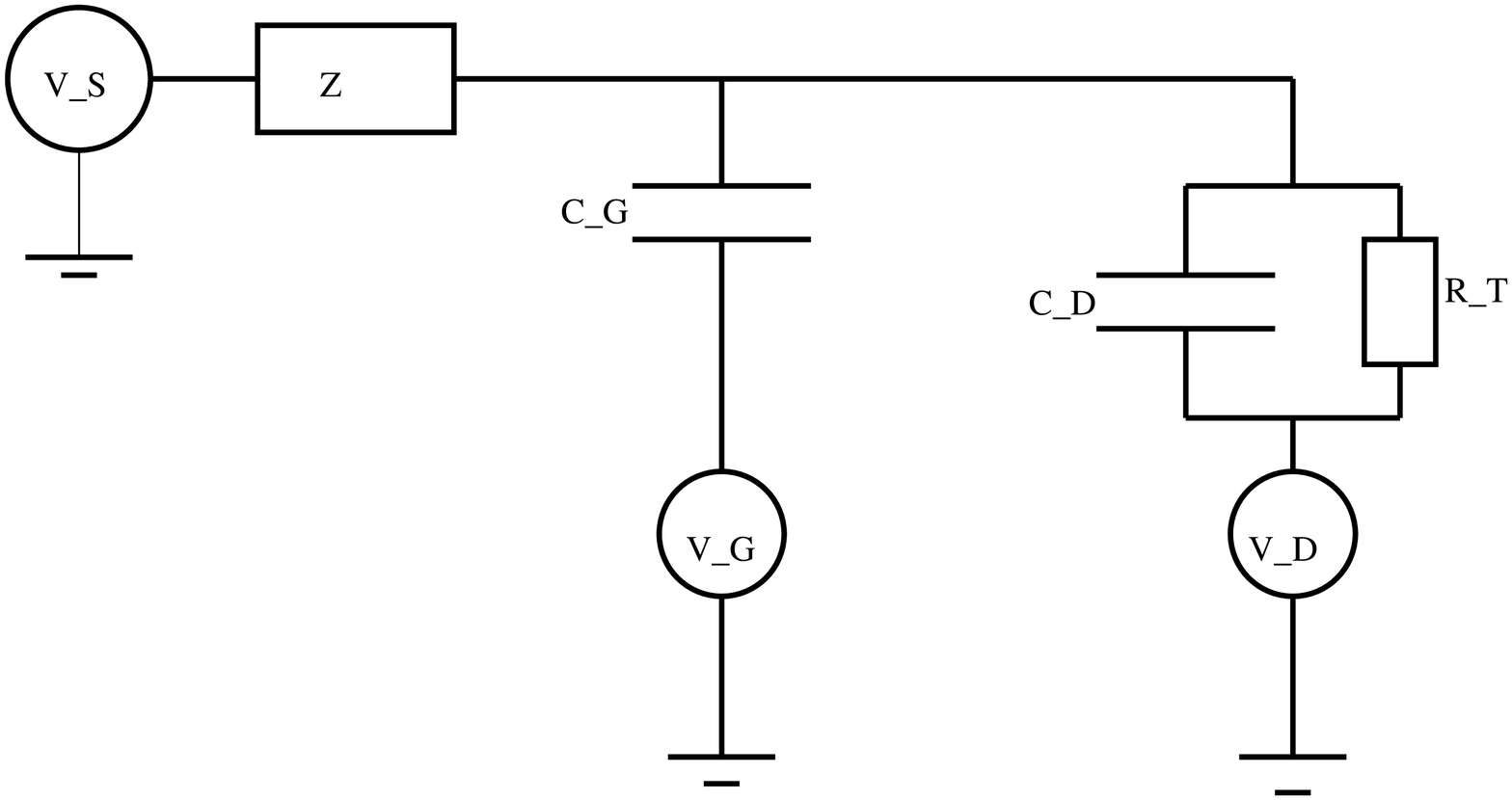}
}
\caption{(a) A schematic picture of the system. One end of a conducting multi-wall carbon nanotube of length $L$, inner diameter $D_{{\rm i}}$, outer diameter $D_{{\rm o}}$ is attached to a source (s) electrode above a step of height $\tilde{h}$ in a silicon dioxide substrate. The other end is free to move. The displacement $x$ of the nanotube tip is measured towards the substrate. A gate (g) and a drain (d) electrode are placed beneath the tube and are used to control the motion of the tip.
(b) An equivalent circuit for the system in Fig.~1(a). The impedance $Z$ is ohmic and describes the tube-source coupling. The tube-gate coupling is purely capacitive with the capacitance $C_{{\rm g}}(x)$ whereas the tube-drain coupling is a tunnel junction with tunneling resistance $R_{{\rm T}}(x)$ and capacitance $C_{{\rm d}}(x)$. $V_{{\rm s,g,d}}$ are the electrostatic potentials on the electrodes and we choose $V_{{\rm d}} = 0$.}
\label{fig:system}
\end{figure*}

A schematic picture of the system is presented in Fig.~\ref{fig:system}. 
A metallic multi-wall carbon nanotube of length $L$ is placed on a terraced silicon dioxide substrate such that its free end is distance $h$ from the drain contact when the tube is horizontal. The fixed end of the nano\-tube is connected to a source electrode, to which the electric contact is purely ohmic with impedance $Z$. Two additional electrodes (gate and drain) are positioned on the lower level. The position along the horizontal axis is labeled $z$. The drain electrode is positioned at $z_{{\rm d}}=L$ and the gate at $z_{{\rm g}} < L$. The ratio $z_{{\rm g}}/z_{{\rm d}} = z_{{\rm r}}$ gives the relative position of the gate. The free end of the tube can move and the electrodes are used to control its motion. The deflection of the nanotube tip from the horizontal position is labeled $x$. Typical length scales of the system are tube length $L = 50 - 150$ nm, and step height $\tilde{h} = 5-10$ nm. By applying a voltage to the gate and/or the drain electrodes, an excess charge $q$ is induced in the tube. This causes an electrostatic force that deflects the tip of the tube towards the drain electrode.  

The tube is modeled using classical continuum elasticity theory, \cite{LL} considering
the CNT to be an elastic cantilever with only the lowest vibrational eigenmode 
excited. Furthermore, we assume that the bending profile of the tube affected by 
an external force $F$ is the same as for free oscillations. Then the potential 
energy of the tube can be expressed in terms of tip deflection $x$ according to 
$V=kx^2/2$ where the effective spring constant is given by 
$k \approx 3EI/L^3$, where
$I = \pi(D_{{\rm o}}^4-D_{{\rm i}}^4)/64$ is the moment of inertia of the cantilever, and $E$ is the Young's modulus, experimentally measured to be of the order of 1 TPa. 
\cite{Treacy, Wong} 
The equation of motion for the tip of the tube is that of a forced  
harmonic oscillator with an effective mass given by 
$m_{{\rm eff}} = k/\Omega^2 \approx 3M_{{\rm CNT}}/(1.875)^2$ where $\Omega$ is the 
lowest vibrational eigenfrequency and $M_{{\rm CNT}}$ is the total mass of 
the CNT-cantilever. 
In addition, a phenomelogical viscous damping 
$\gamma_{\rm d} \dot{x}$ is introduced into the equation of motion,  
$\gamma_{\rm d}$ is estimated from Q-factors from experimental data. \cite{poncharal} 
Finally we arrive at the equation of motion
\begin{equation}
m_{{\rm eff}}\ddot{x} = - kx - \gamma_{\rm d} \dot{x} + F_{{\rm c}} + F_{{\rm vdW}} + F_{{\rm sr}},
\end{equation}
where $F_{\rm c}$ is the capacitive force, $F_{{\rm vdW}}$ is the van der Waals force, and $F_{{\rm sr}}$ are the short range forces due to wave function overlap. These different forces will be discussed in more detail in the next section. The charge dynamics in the tube is governed by
\begin{equation}
\begin{split}
Z\dot{q} =& -\frac{q}{C_{{\rm g}}(x) + C_{{\rm d}}(x)} 
 + V_{{\rm s}} \\ &- V_{{\rm g}}\frac{C_{{\rm g}}(x)}{C_{{\rm g}}(x) + C_{{\rm d}}(x)} - Z I_{{\rm sd}},  
\end{split}
\end{equation}
where $V_{\rm s}$ and $V_{\rm g}$ are the electrostatic potentials on the source and the gate, respectively, $Z$ is the source junction impedance, $C_{{\rm g}}$ and $C_{{\rm d}}$ are the capacitances between the tube and the gate and the drain respectively, and $I_{{\rm sd}}$ is a stochastic tunneling current between the tube and the drain electrode. Due to the much larger tube-gate separation tunneling to the gate is neglected. We choose $V_{{\rm d}} = 0$.

The tunneling rates of the stochastic tunneling current are 
calculated using the theory of Coulomb blockade in the presence of 
an electromagnetic environment. 
Following Ingold and Nazarov, \cite{ingoldnazarov}
assuming an ohmic environment and zero temperature, the tunneling rate 
 $\Gamma$ for a given deflection $x$ is determined by  
\begin{equation}\label{eq:tunrate1}
\Gamma(x) = \frac{1}{e^2 R_{{\rm T}}(x)} \int_0^{eV}(eV - E) P(E) dE,
\end{equation}
where $R_T$ is the tunneling resistance and $V$ is the electrostatic 
potential difference
between the tube and the drain contact ($V \approx V_{\rm s}$). 
Furthermore, $P(E)$ is the probability 
for energy exchange between the
tunneling electron and the environment and is self-consistently
determined from 
\begin{equation}\label{eq:tunrate2}
E P(E) = \frac{2 Z}{R_{{\rm K}}} \int_0^E dE' \bigg{[}1+\bigg{(}\frac{\pi
Z}{R_K}\bigg{)}^2\bigg{(}\frac{E-E'}{E_{{\rm c}}(x)} \bigg{)}^2 \bigg{]}^{-1} P(E'), 
\end{equation}
where $E_c(x) = e^2/[2C_{\Sigma}(x)]$, where $C_{\Sigma}(x) = C_{{\rm d}} + C_{{\rm g}}$ is the total capacitance between the tube and the contacts and $R_{{\rm K}} = 2\pi \hbar /e^2 \approx 25.8$ k$\Omega$ is the von
Klitzing constant. The tunneling resistance is approximated by 
\begin{equation}\label{eq:tunres}
R_{{\rm T}} =  R_0 e^{(h-x)/\lambda}, 
\end{equation}
where $R_0$ is the tunneling resistance when the tube is in mechanical contact with the electrode, $\lambda$ is the material dependent tunneling length of the order of 0.5~\AA~and $h$ is the tip-contact separation at zero deflection. Note that $h < \tilde{h}$ since the contact is deposited on top of the substrate.

\section{FORCES}
\subsection{Electrostatic Force}

A numerical solution to the Poisson equation using a finite element method \cite{torgny} suggests that we can approximate the capacitances of the junctions with a parallel plate capacitor model with an offset. The capacitances between the tube and the gate electrode $C_{{\rm g}}$ and the tube and the drain electrode $C_{{\rm d}}$ are given by \footnote{This macroscopic treatment can also be regarded as a first order Taylor expansion of $1/C(x)$.}
\begin{eqnarray}
C_{{\rm d}}(x) = \frac{C_0}{1-\frac{x}{h}(1-C_0/C_{h})}, \label{eq:dcap} \\
C_{{\rm g}}(x) = \frac{2 C_0}{1-\kappa \frac{x}{h}(1-C_0/C_{h})}, \label{eq:gcap}
\end{eqnarray}
where $C_0$ is the drain junction capacitance at zero tube deflection, $C_{h}$ is the drain junction capacitance when the tube is in mechanical contact with the drain electrode, and $\kappa = 0.5 (3 z_r^2 -z_r^3)$ takes into account that the deflection at the tube above the gate is smaller than at the tip. The value of $C_h$ is estimated from experimental data \cite{tarkiainen} and $C_0$ is calculated for parallel plate capacitors. The resulting capacitive forces are given by 
\begin{equation}\label{eq:capforce}
F_{{\rm c}} = -\nabla \left( \frac{q_{{\rm d}}^2}{2 C_{{\rm d}}} + \frac{q_{{\rm g}}^2}{2 C_{{\rm g}}}\right),
\end{equation}
where $q_{{\rm d, g}}$ are the charges on the drain and gate capacitors respectively,
 $q_{{\rm d}} + q_{{\rm g}} = q$, and $q$ is assumed to be constant during an 
infinitesimal displacement of the tube. 

\subsection{Van der Waals force}
The origin of the van der Waals (vdW) force is correlation of fluctuating higher order electrostatic moments between two charge distributions. The force is attractive and the interaction energy varies with separation as the inverse sixth power of separation in the range of separations relevant to the nanorelay. \cite{bruch}   
The total vdW-energy between the tube and the substrate plus contacts is calculated in a continuum limit. \cite{Dequesnes} The pairwise sum over the interaction between the individual atoms is transformed into a six-dimensional integral over the volumes of the interacting bodies. In this approximation we can write the vdW-force as
\begin{equation}
\begin{split}
F_{{\rm vdW}} &= - \nabla E_{{\rm vdW}} \\ &= - \nabla \int_{V_{1}} \int_{V_{2}} \, dV_1 dV_2 n_1 \frac{-C_6}{|\mathbf{r_1} - \mathbf{r_2}|^6} n_2,   
\end{split}
\label{eq:vdWforce}
\end{equation}
where $V_1$ is the volume of the tube, $V_2$ is the volume of the contacts and substrate, and $n_1(\mathbf{r_1})$ and $n_2(\mathbf{r_2})$ are the atom densities in the two bodies. The values of the interaction parameter $C_6$ depend on the species of the interacting atoms and is usually stated in terms of the Hamaker constant $A = \pi^2 n_1 n_2 C_6$. The Hamaker constant for interaction between graphite-silicon dioxide is given by $A = 0.07$ aJ and for graphite-metal the corresponding value is 0.6 aJ. \cite{goodman} 

Evaluating the integral in Eq.~(\ref{eq:vdWforce}) is time-consuming for 
an arbitrary geometry. To simplify this, we separate $E_{{\rm vdW}}$ in two parts, one contribution from tube-electrodes interaction and one from tube-substrate interactions. The substrate is approximated by a semi-infinite plane and the drain electrode is assumed to be cylindrical and infinitely long. Furthermore, we assume that the interaction between the tube and the gate electrode is much smaller than the  tube-drain interaction due to the much larger separation.
Assuming the MWNT to consist of individual CNTs separated by 3.4 \AA, we transform the integral over one-dimension to sum over all individual CNTs in the MWNT. Many of the integrals in (\ref{eq:vdWforce}) can be carried out analytically, leaving either one (for tube-substrate interaction) or three (for tube-electrode interaction) to be evaluated numerically.

\subsection{Short Range Forces}

The attractive part of the short range force $F_{{\rm att}}$ originates from a coupling between the tube and the contact when the electronic wave functions overlap. It can be roughly related to the tunneling conductance $G_{\rm T} = 1/R_{\rm T}$ between the tube and the contact as \cite{chen} 
\begin{equation}
F_{{\rm att}} \propto \sqrt{G}.
\end{equation}
A more detailed analysis of the this force is rather complicated but 
does not change the qualitative features of the system's behavior. 
\cite{batra} 
When the core-core separation of the of the atoms in the tube and in the tip becomes small, however, the wave function overlap increases and we expect a strong repulsion due to the Pauli principle. Taking this repulsive part $F_{{\rm rep}}$  to be exponential we model the attractive and the repulsive short range forces by a Morse curve\cite{slater}
\begin{equation}
\begin{split}
F_{{\rm sr}} &= F_{{\rm att}} + F_{{\rm rep}} \\  &= \frac{f}{2\lambda} 
\sqrt{\epsilon_{{\rm C}} \epsilon_{{\rm T}} R_{{\rm K}} G_0} 
(e^{-\frac{1}{2\lambda}(h-x)} - e^{-\frac{1}{\lambda}(h-x)}),
\end{split}
\end{equation}
where $f$ is a dimensionless shape factor of the order of one, $\epsilon_{{\rm T }}$ and $\epsilon_{{\rm C}}$ are the valence band widths of the tube and the drain electrodes, respectively, and $G_0=1/R_0$ is the conductance at contact. We define the location of the surface by $F_{{\rm sr}} = 0$. The core-core separation between the atoms in the tube and the drain electrode when the tube is in mechanical contact with the electrode  is assumed to be 3 \AA.  This offset is introduced to prevent the unphysical divergence in the vdW-energy at zero tube-drain electrode separation. 

\begin{figure}
\centering
\psfrag{xlabel}[t]{\small{$x$ [$h$]}}
\psfrag{ylabel}[b]{\small{Force [nN]}}  
\includegraphics[width=3.3in]{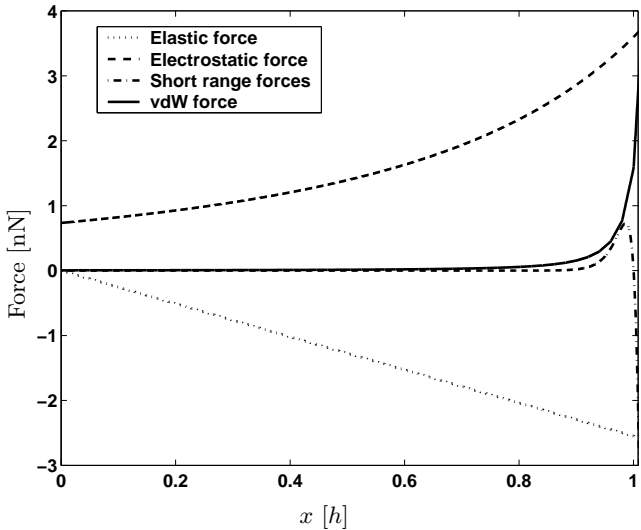}
\caption{The magnitude of the different forces in the system with parameters given 
in chapter V. The voltages $V_{{\rm g}} = 7.5$ V and $V_{{\rm g}} = 0.01$ V were used. 
Close to the contact located at $x=h$ the forces are of the same order while 
electrostatic and elastic forces dominate far from the surface.}
\label{fig:forces}
\end{figure} 

\begin{figure}
\psfrag{ylabel}[b]{\small{$x$ [$h$]}}
\psfrag{xlabel}[t]{\small{$V_{{\rm g}}$ [V]}}
\centering
\includegraphics[width=3.3in]{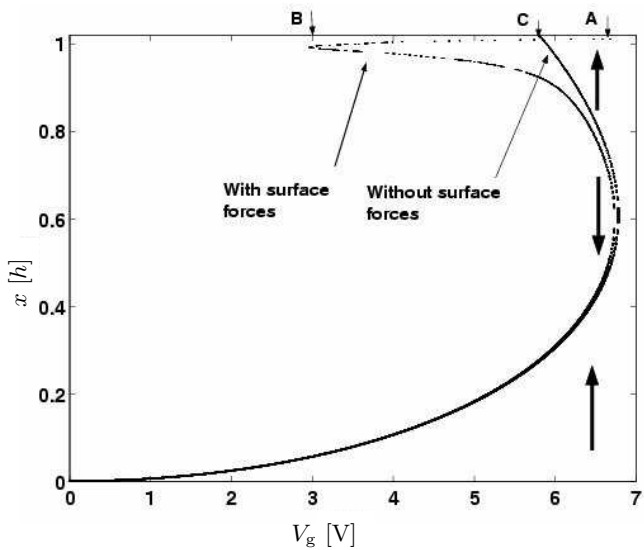}
\caption{Stability diagram with and without surface forces for the system 
with parameters given in Chapter V. The curve shows the
positions of zero net force on the tube (or local equilibria) as functions of gate 
voltage (at constant $V_{{\rm s}} = 0.01$ V) and deflection $x$ (in units of $h$). 
The large arrows shows the direction of the force on each side of 
the curves, 
indicating one local equilibrium to be unstable. The required voltage for pulling
the tube to the surface (``pull-in voltage'') is given by A 
($\approx 6.73$ V). In agreement with 
reference \onlinecite{Dequesnes} this voltage is not significantly 
affected by surface forces. 
A tube at the surface will not leave the surface until the voltage is lower than 
the ``release voltage'', B and C in the figure. Note that A $>$ B, C which 
indicate a hysteretic behavior in the IV$_{{\rm g}}$--characteristics, a feature 
significantly enhanced by surface forces.}
\label{fig:stability}
\end{figure}

\section{Phonon excitation}
The only dissipative mechanism in the equation of motion thus far is a viscous 
damping. Several other dissipative mechanisms are present in a real system.
One effect that is important for the device performance is the 
possibility of phonon excitation
in the drain contact when the tube bounces off the surface of the drain electrode. 
This surface dissipation is important for the switching behavior. 

\subsection{Phonon model}    
To estimate the dissipation associated with 
phonon emission we use a simplified classical one dimensional model 
of the atomic structure. The atomic lattice
is modeled as a collection of atoms, each connected to the other
through a harmonic potential, characterized by a spring constant $K$. 
Assuming linear dispersion relation we obtain the spring constant \cite{AM} 
\begin{equation}
K = \frac{Mc^2}{a^2},
\label{eq:springk}
\end{equation}
where $M$ is the mass of the atoms in the chain, $a$ is the equilibrium 
separation between atoms, and $c$ is the speed of sound in the 
electrode material. We consider the drain electrode to be made of gold, 
and thus, we use the 
lattice constant of gold $a = 407.8$ \AA, the mass of one gold atom 
$M = 197$ amu, and the sound velocity in gold $c = 2000$ m/s.
\begin{figure}
\psfrag{xlabel}[t]{$v_{{\rm in}}$ [m/s]}
\psfrag{ylabel}[b]{$E_{{\rm out}}$ [$E_{{\rm in}}$]}
\centering
\includegraphics[width=3.3in]{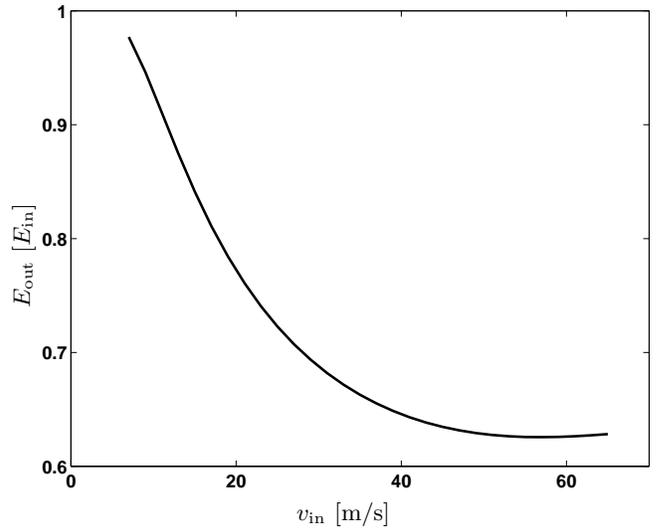}
\caption{The remaining kinetic energy of the tube in units of incident kinetic 
energy as a function of incident velocity. Increasing the incident velocity 
increases the relative dissipation. The nanotube has an effective mass of 
$2.1 \cdot 10^{-21}$ kg corresponding to the effective mass for the 
typical system described in chapter V.}
\label{fig:loss}
\end{figure} 

On the atomic scale the nanotube is a very large object and, thus, the tube 
will interact with the surface over an extended region while bouncing. 
We estimate that the impact area of the 
nanotube corresponds to about 50 gold atoms (an impact area of the
order of $2$ nm$^2$) in the contact. Instead of considering the motion
of each atom in the lattice, we consider the motion of layers of atoms, 
each layer made up of 50 atoms. The nanotube, characterized by the 
effective mass $m_{{\rm eff}}$, impinges on the model lattice with 
an initial velocity $v_{{\rm in}}$, interacts with the topmost layer
through a repulsive force $F_{{\rm rep}}$ and bounces. We use a total 
number of 100 layers, a number chosen such that the boundary conditions
at the last layer have only minor importance.  
The motion of the lattice layers and the nanotube is governed by
classical dynamics and the resulting one-dimensional problem is  
solved numerically. 

The one-dimensional model we use does not, \emph{e.g.}, take 
into account phonons with different polarizations, 
hence, the model most likely underestimates surface dissipation.

\section{Results}

\subsection{Effects of Surface Forces}

The surface forces introduce constraints on the design parameters. 
For the operation of the relay it is necessary that 
\begin{eqnarray}
-kx + F_{\rm c} + F_{{\rm vdW}} + F_{{\rm sr}} &>& 0 \label{eq:pullin}\\ 
-kx + F_{{\rm vdW}} + F_{{\rm sr}} &<& 0 \label{eq:release}
\end{eqnarray}
in order to both enable pulling the tube to the contact (\ref{eq:pullin}) and for the tube to release when the voltage is turned off (\ref{eq:release}). In particular, the latter condition is hard to fulfill -- a manifestation of the ubiquitous stiction problem in nano-science. 

We investigate the behavior of a typical system  
where Eqs.~(\ref{eq:pullin}) and ~(\ref{eq:release}) are satisfied. The geometrical parameters of this system are tube length $L = 120$ nm, inner tube diameter $D_{{\rm i}} = 2$ nm, outer tube diameter $D_{{\rm o}} = 8.8$ nm, tube contact separation at zero deflection $h = 5$ nm, relative position of the gate $z_{{\rm r}} = 0.7$, and Young's modulus $E = 1$ TPa. With these geometrical parameters the effective mass is $m_{{\rm eff}} \approx 2.1 \cdot 10^{-21}$ kg. 
Other parameters are source junction impedance $Z = 8$~k$\Omega$, quality factor of the viscous damping $Q = 250$, capacitance at zero deflection $C_0 = 1.3 \cdot 10^{-19}$ F, and capacitance at contact $C_h = 5 \cdot 10^{-18}$ F.    

The magnitudes of the forces in a typical system are depicted in Fig.~\ref{fig:forces} as a function of the deflection of the tube. When the tube tip is near the drain electrode, all forces are comparable, while far from contact the electrostatic and mechanical forces dominate. 


\begin{figure}
\centering
\psfrag{ylabel}[b]{\small{$I_{{\rm sd}}$ [arb. units]}}
\psfrag{xlabel}[t]{\small{t [s]}}
\includegraphics[width=3.3in]{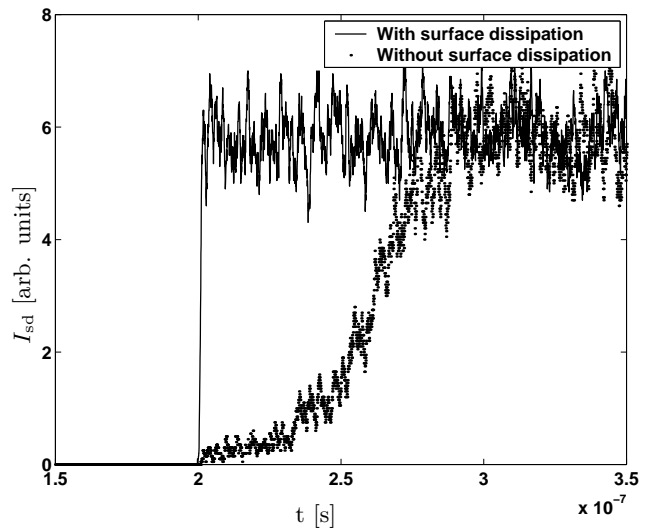}
\caption{Step response (gate voltage step ($0 \rightarrow -7.5 $ V) applied at $t=2 \mu$s) of the system with parameters given in chapter V without phonon dissipation (elastic bounce) and with surface dissipation (inelastic bounce). The dissipative mechanism reduces the total time of switching by nearly two orders of magnitude. The noise reflects the stochastic 
nature of the tunneling between the tube and the drain electrode (shot noise).}
\label{fig:comp}
\end{figure}

\begin{figure}
\psfrag{xlabel}[t]{\small{$V_{{\rm g}}$ [V]}}
\psfrag{ylabel}[b]{\small{$I_{{\rm sd}}$ [nA]}}
\centering
\includegraphics[width=3.3in]{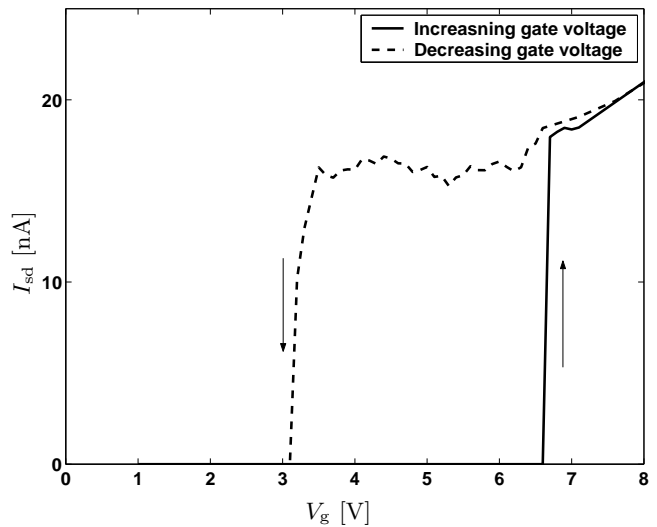}
\caption{The IV$_{{\rm g}}$--characteristics for a typical system  
with the parameters given in Chapter V has a large hysteresis loop. The arrows 
indicate up- or down sweep of $V_{{\rm g}}$ (at constant $V_{{\rm s}}= 0.01$ V). 
The system switches to the conducting state at $V_{{\rm g}} \approx 6.7$ V which 
agrees well with the expected value deduced from Fig.~\ref{fig:stability}. The 
reverse transition when lowering the gate voltage takes place at $V_{{\rm g}} 
\approx 3.1$ V which is slightly larger than the expected value. This small 
discrepancy is due to small vibrations of the tube and non-adiabatic voltage sweep.}
\label{fig:hysteres}
\end{figure}

\begin{figure}[htb]
\centering
\psfrag{xlabel}[t]{\footnotesize{t [s]}}
\psfrag{ylabel}[b]{\footnotesize{$I_{{\rm sd}}$ [arb. units]}}
\subfigure[$T_{{\rm w}} = 0.015$ ns. The system stays
in the conducting 1-state.] 
{
\label{fig:ejnollasurface}
\includegraphics[width=1.5in]{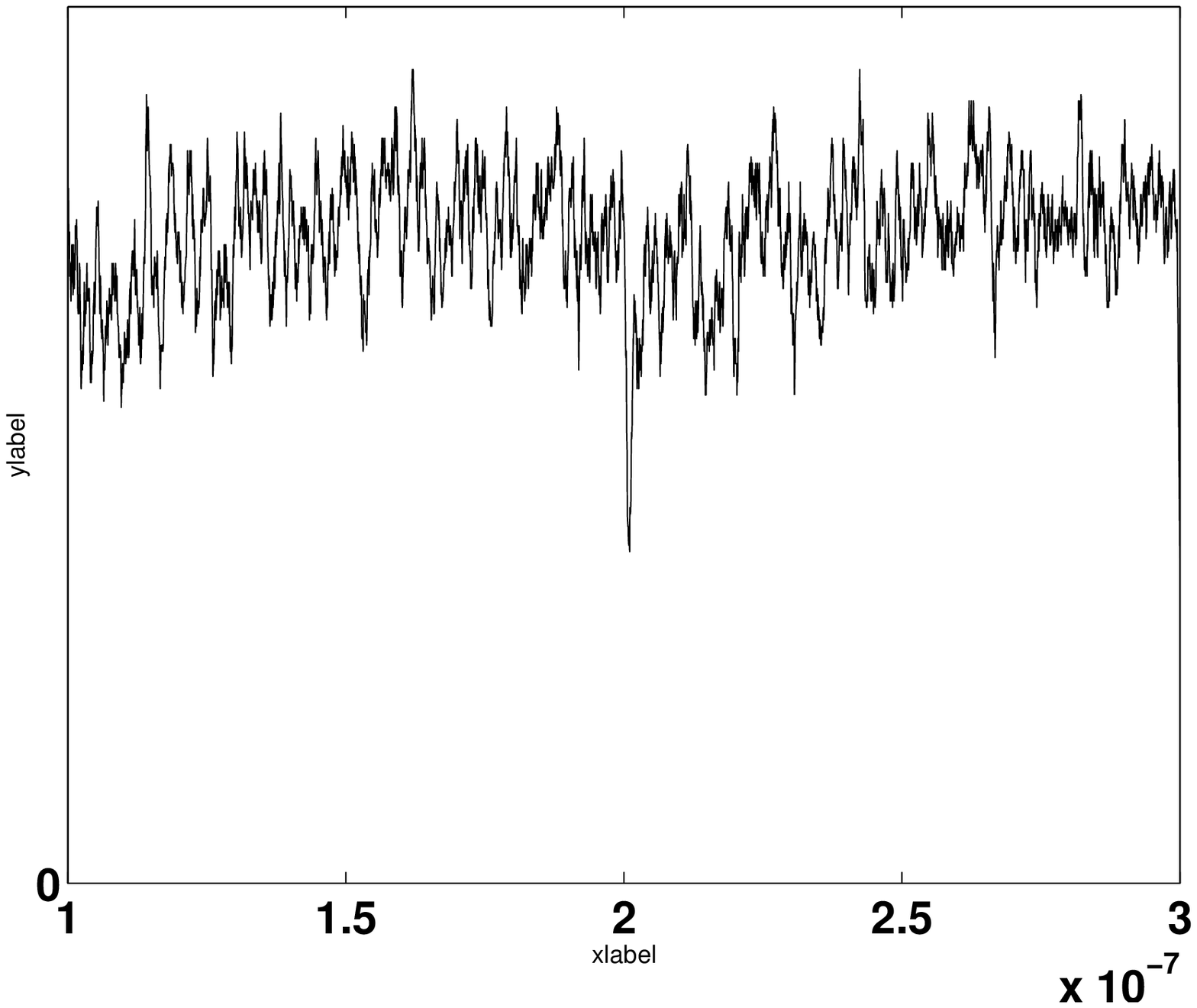}
}
\hspace{0cm}
\subfigure[$T_{{\rm w}} = 0.020$ ns. The system switches to
the non-conducting 0-state.] 
{
\label{fig:nollasurface}
\includegraphics[width=1.5in]{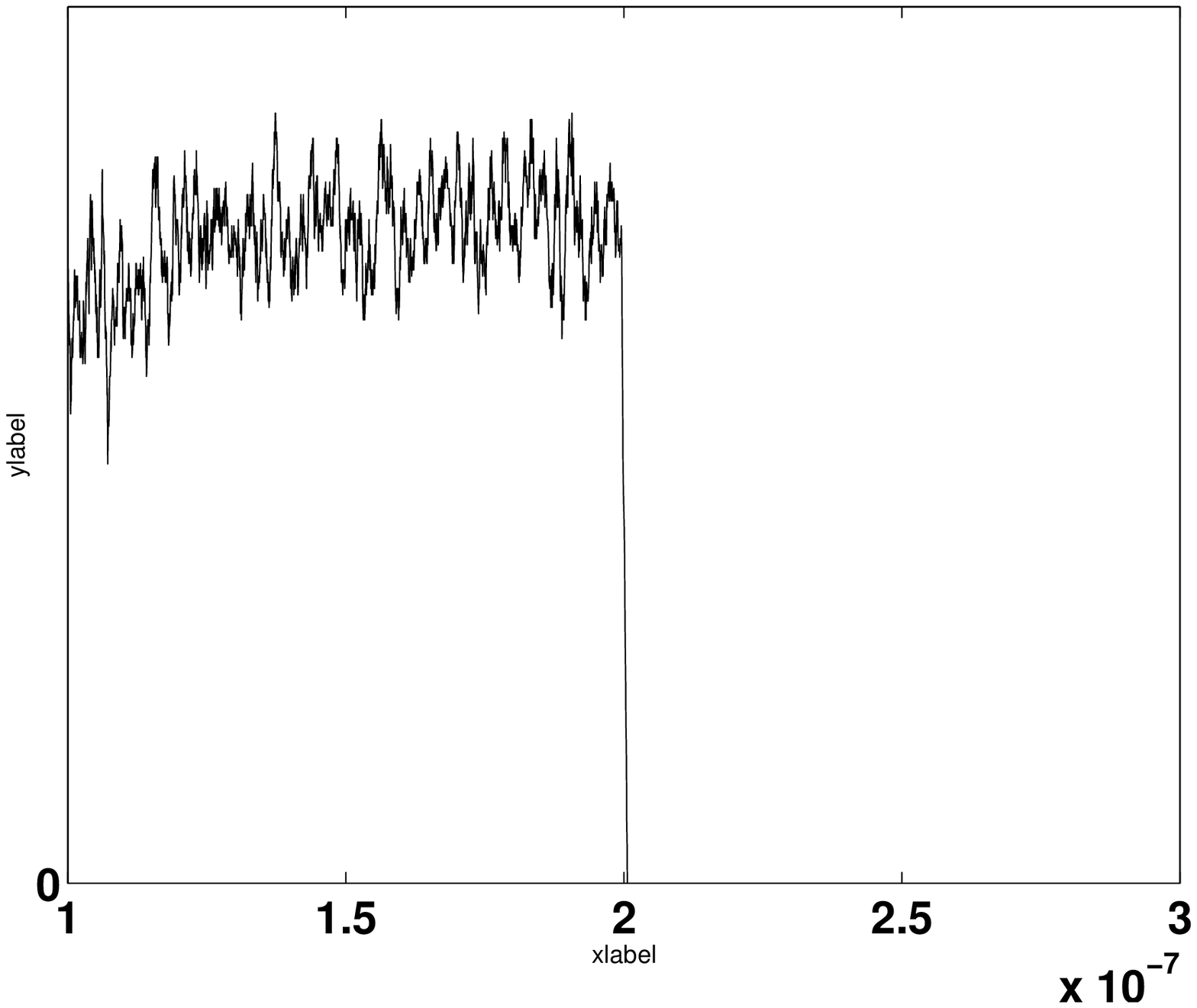}
}
\\
\subfigure[$T_{{\rm w}} = 0.7$ ns. The system stays
in the non-conducting 0-state.] 
{
\label{fig:ejettasurface}
\includegraphics[width=1.5in]{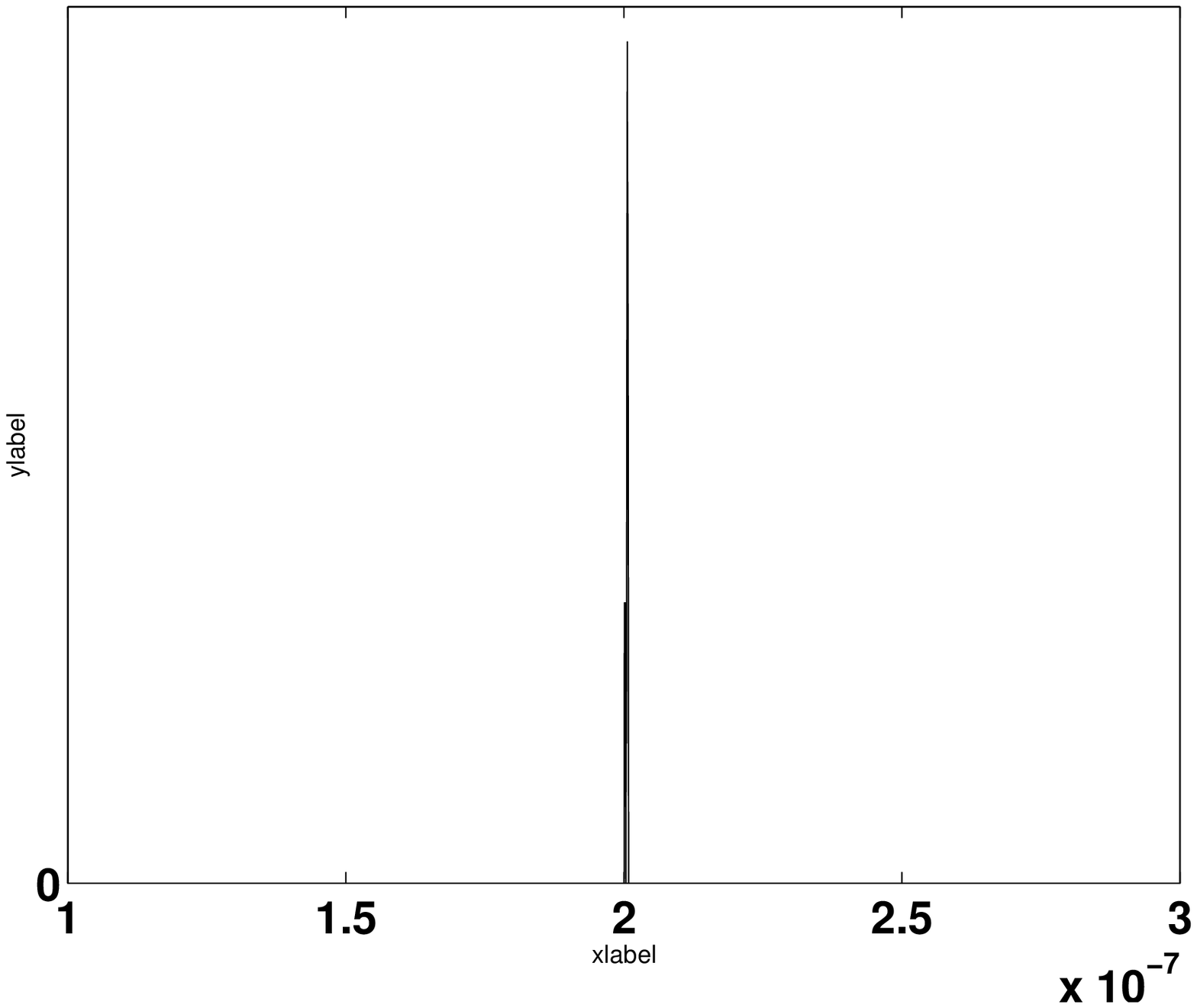}
}
\hspace{0cm}
\subfigure[$T_{{\rm w}} = 0.8$ ns.  The system switches
the conducting 1-state.] 
{
\label{fig:ettasurface}
\includegraphics[width=1.5in]{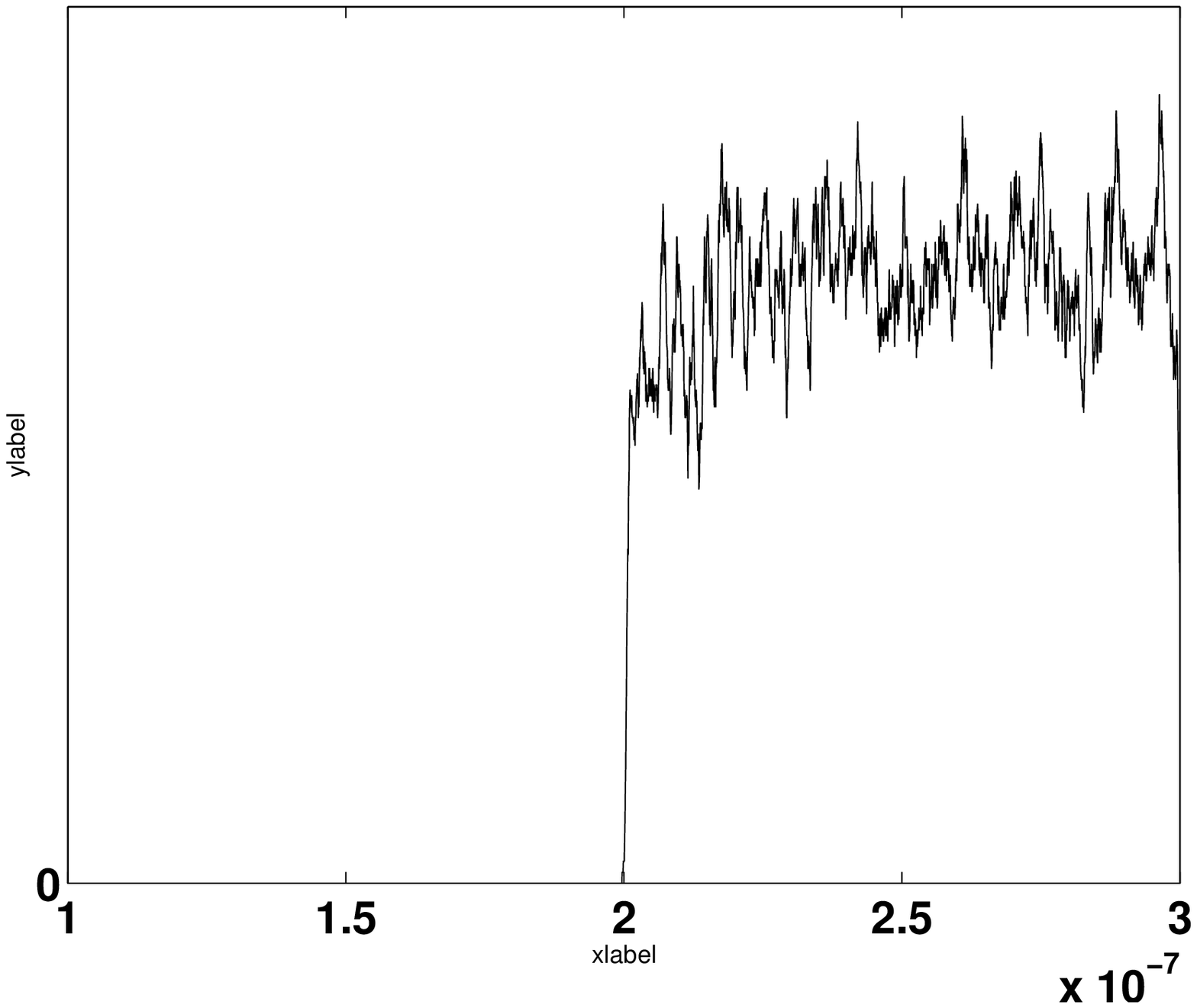}
}
\caption{The system's response for gate voltage pulses of duration $T_{{\rm w}}$ applied at $t= 0.2$ $\mu$s when biased within the hysteresis loop ($V_{{\rm g}} = 5$ V). The parameters of the system are given in chapter V. (a,b) low voltage pulse $V_{{\rm g}}(t) = 0$ V to write a `0'; (c,d) high voltage pulse $V_{{\rm g}}(t) = 7.5$ V to write a `1'. Note the asymmetry between the write times of a `0' and a `1'. }
\label{fig:writetime} 
\end{figure}

The effects of the surface forces can be visualized by means of a stability diagram 
which shows the positions of zero net force (local equilibria) on the cantilever 
as a function of gate voltage and deflection. A typical stability curve for the 
system is given in Fig.~\ref{fig:stability}. 
A stability diagram with more than one local equilibrium for a specific voltage 
results in a hysteretic behavior in the IV$_g$--characteristics (more than one 
position of zero net force means that there actually are three such positions, 
one of which is not stable). This is because the net force is positive to the 
right of the curve and negative to the left of the curve, so that, when 
lowering the voltage for a tube in contact with the drain electrode, it will 
not release until there is no stable position at the surface. 
Fig.~\ref{fig:stability} shows that this effect is significantly increased 
by the surface forces. 

The ``sticking problem'' corresponds to a stable nanotube position for 
zero voltage at the surface, \emph{i.e.} failure to satisfy ~(\ref{eq:release}). 
This problem can be alleviated \emph{e.g.} by choosing stiffer or shorter tubes. 
This in turn leads to higher voltages needed to satisfy equation ~(\ref{eq:pullin}). 
With high voltages and small distances, the electric field strengths can be very 
high. Under these conditions field emission may be important. Simulations show, 
however, that design parameters can be chosen so that this has a negligible 
effect. \cite{sven}

\subsection{Effects of phonon dissipation}
As the incident velocity of the tube increases a larger percentage 
of energy is transferred 
from the tube to the substrate. 
For our particular set of simulation parameters 
we get a loss function according to Fig.~\ref{fig:loss}.
For high incident velocities we get high dissipation 
whereas for low velocities the relative dissipation gets smaller. This 
property makes relaxation towards the surface fast even for high 
initial velocities.  

The main effect of exciting phonons in the drain contact is to change the 
switching dynamics of the relay. Without this dissipative mechanism, 
relaxation of the tube towards a stationary current carrying state
is slow due to 
the fact that the tube bounces off the contact surface many times before 
coming to a rest.  This 
time scale is significantly decreased when phonon dissipation is included in 
the model. A comparison between a simulation showing the response to 
an applied step gate voltage without dissipation and an equivalent 
simulation with surface dissipation, depicted 
in Fig.~\ref{fig:comp}, show that the time scale of switching is 
decreased by two orders of magnitude. The reverse transition is not 
affected by including phonon dissipation in the model. 

\subsection{Memory Element Application}
We have performed numerical simulations of the system described above and  
the resulting I$V_g$--characteristics at a constant $V_{{\rm s}} = 0.01$ V are 
shown in Fig.~\ref{fig:hysteres}. 
The hysteresis is large, which makes a memory element a promising application of the device. In such a memory element the conducting state can be defined as a logical 1 and the non-conducting state as a logical 0. The system is biased at a gate voltage chosen to lie within the hysteresis loop where the system is bistable. The element is written by applying a high/low voltage for a specific write time, $T_{{\rm w}}$. The low voltage is typically zero and high voltage is chosen such that the condition ~(\ref{eq:pullin}) is fulfilled. The write times must be long enough to ensure that the correct state is reached, and depend on the chosen write voltages. 

We have estimated the write times of the memory element by numerical simulations. 
Such simulations for the chosen set of parameters are depicted 
in Fig.~\ref{fig:writetime}. 
The write time for $0 \rightarrow 1$ is approximately $0.8$ ns and 
for $1 \rightarrow 0$ the write time is about $0.02$ ns. The difference between
these time scales can be explained by comparing the two write processes: When 
the tube moves from a stationary 1, the tunneling resistance increases 
exponentially with tube contact separation and the current will stop soon 
after the voltage is switched off. In the $0 \rightarrow 1$ transition the 
tube bends relatively quickly to the drain electrode, but tends to bounce 
off the surface. The kinetic energy of the mechanical motion must be dissipated 
before the tube relaxes toward the stationary 1-state.  This relaxation is 
enhanced due to phonon dissipation in the drain, but we cannot expect it
to remove the asymmetry completely. The time scale corresponding 
to the $0 \rightarrow 1$ transition sets the limit for maximum operating frequency,
approximately 1 GHz for the parameters we have used. 

\section{Conclusions}
We have incorporated short range and vdW-forces into a model of a three terminal nanorelay.  The main effect of these forces is to increase the hysteresis in the $IV_{{\rm g}}$--characteristics making a memory element a promising application of the relay. We have investigated the switching characteristics for such a memory element and conclude that the write dynamics is limited by the $0 \rightarrow 1$ transition. The transition dynamics of the $0 \rightarrow 1$ transition depends to a great extent on the dissipative surface processes when the tube bounces, which reduce the transition time by two orders of magnitude, whereas the $1 \rightarrow 0$ transition is unaffected by such processes.

\begin{acknowledgments}
We would like to thank the Swedish Foundation for Strategic Research for funding this project through the CARAMEL (Carbon Allotropes for Microelectronics) consortium and the framework program on CMOS-integrated carbon-based nanoelectromechanical systems. One of the authors (T.N.) acknowledges financial support from the Swedish research council (VR).
\end{acknowledgments}

\end{document}